\documentclass[%
reprint,
superscriptaddress,
 amsmath,amssymb,
 aps,
longbibliography,
floatfix,
]{revtex4-2}

\usepackage[dvipsnames,svgnames,x11names,hyperref]{xcolor}
\usepackage{siunitx}
\usepackage{tabularx}
\usepackage{color}
\usepackage{graphicx}
\usepackage{dcolumn}
\usepackage{bm}
\usepackage{hyperref}
\usepackage[mathlines]{lineno}
\usepackage{mathrsfs}
\usepackage{braket}
\usepackage{placeins} 
\usepackage{float} 
\usepackage{color}
\usepackage{esdiff} 

\hypersetup{colorlinks=true, 
	linkcolor={blue!75!black!80!yellow},
	citecolor={blue!75!black!80!yellow}, 
	urlcolor=black 
	}
\makeatletter \renewcommand\@make@capt@title[2]{%
\@ifx@empty\float@link{\@firstofone}{\expandafter\href\expandafter{\float@link}}%
\sffamily{\textbf{#1}}\@caption@fignum@sep#2 }

\begin{document}

\title{Entangled photons from composite cascade emitters}

\author{Derek S. Wang}
\altaffiliation{Contributed equally to this work}
\affiliation{Harvard John A. Paulson School of Engineering and Applied Sciences, Harvard University, Cambridge, MA 02138, USA}

\author{\.Inci Anal{\i}}
\altaffiliation{Contributed equally to this work}
\affiliation{Harvey Mudd College, Claremont, CA 91711, USA}

\author{Susanne F. Yelin}
\email{syelin@physics.harvard.edu}
\affiliation{Department of Physics, Harvard University, Cambridge, MA 02138, USA}

\begin{abstract}
\noindent Entangled photons are crucial for quantum technologies, but generating arbitrary entangled photon states deterministically, efficiently, and with high fidelity remains a challenge. Here, we demonstrate how hybridization and dipole-dipole interactions---potentially simultaneously available in colloidal quantum dots and molecular aggregates---leveraged in conjunction can couple simple, well understood emitters into composite emitters with flexible control over the level structure. We show that cascade decay through carefully designed level structures can result in emission of frequency-entangled photons with Bell states and three-photon GHZ states as example cases. These results pave the way toward rational design of quantum optical emitters of arbitrarily entangled photons.
\end{abstract}
\date{\today}

\maketitle

\section{Introduction} \label{sec:intro}

Entanglement, especially among photonic qubits, is not only useful for testing the limits of quantum theory \cite{Einstein1935, Paul1964, Freedman1972, Shalm2015, Pan2012}, but also a valuable resource in photonic quantum technologies, such as quantum computers and quantum networks \cite{Pan2012, Nielsen2011, Wehner2018, Hillery1999, Nielsen2006}. For instance, Bell states, or maximally entangled photon pairs, are necessary for quantum teleportation, the fundamental mechanism by which quantum repeaters send quantum information over long distances \cite{Wehner2018}; three-qubit maximally entangled GHZ states are useful for quantum cryptography and secret sharing \cite{Hillery1999}; and cluster states, or highly entangled states of many qubits, underlie measurement-based quantum computing that is formally equivalent to more traditional quantum circuit-based models but requires only easy-to-implement single-qubit gates upon successful creation of a cluster state \cite{Nielsen2006}.

Despite the ubiquitous need for entangled photons in quantum technologies, producing them with high fidelity, quickly, and deterministically---even just pairs of photons---remains a challenge. Relatively successful approaches for producing pairs of entangled photons include spontaneous parametric down-conversion \cite{Kwiat2001, Burnham1970, Howell2004, Horn2012, anwar2020entangled} or four-wave mixing \cite{takesue2004, Lu2019}, but the number of photon pairs generated follows a Poissonian distribution \cite{Waks2004}, rendering both the pair generation efficiency and rate too slow for scalable quantum systems \cite{Pan2012}. Another approach uses semiconductor quantum dots to deterministically emit entangled photon pairs \textit{via} biexciton decay cascade \cite{Akopian2006, Muller2009, Chen2016, Orieux2017, Huber2018, Zeeshan2019, Liu2019, Fognini2019, ahmadi2020toward} with high fidelity and emission efficiency. Generalizing this approach to produce higher-order, multi-photon entangled states with fine control over the entanglement basis remains difficult, however, driving long-standing and active research into alternative approaches \cite{Zukowski1998, Zwierz2009, Gimeno-Segovia2019} that so far require complex, active control over photon emitters.

Given the intuitive simplicity of cascade decay from multiply excited emitter states to produce entangled photons, we consider how to engineer their level diagrams for generation of arbitrarily entangled photons. In particular, we turn to recent work in constructing composite emitters from well understood single-emitter building blocks. One method of coupling together single emitters is through the hybridization (or charge transfer) interaction, where wave function overlap between states of individual emitters placed close together results in a composite emitter with additional excited states with shifted energies. This effect was computationally demonstrated for defects in solid-state materials \cite{Wang2020Hybridized}, where the excitation energy of a defect center in hexagonal boron nitride predicted to be a single-photon emitter could be shifted across the entire visible range, as well as in fused colloidal quantum dots \cite{Cui2019}. Another method of coupling together single emitters is through the transition dipole-dipole interaction that has been used to create composite emitters capable of emission of entangled photon pairs \cite{Wang2020Entangled}, implementation of multi-photon quantum logic gates \cite{Lukin2000, Dai2020}, and superradiance \cite{Philbin2021}. Often, emitters are assumed to only interact through either the shorter-range hybridization or longer-range dipole-dipole coupling, but not both. However, at intermediate distances between emitters, both types of interaction can be simultaneously relevant, such as in
moir\'e excitons \cite{Yu2017}, molecular aggregates \cite{Hestand2018}, and quantum dots \cite{Cui2019}. Are there, then, benefits to using both the hybridization and dipole-dipole coupling interactions?

In this study, we demonstrate how to design the level structures of composite quantum emitters for deterministic emission of entangled photons \textit{via} cascade decay by coupling together single emitters through both the hybridization and the dipole-dipole coupling interactions. We show that the two types of interactions lead to qualitatively different types of state mixing and that this difference can be leveraged to generate a level structure amenable to emission of two-photon Bell states from just two two-level emitters. We also show that this approach can be generalized to higher-order entangled states by designing a system consisting of three emitters that is capable of emitting three-photon GHZ states with simultaneous efficiency $\eta$ and fidelity $\mathcal{F}$ as high as 90\%. We anticipate these results will motivate research into designing composite emitters for emission of entangled photons from real emitters, especially molecular aggregates and colloidal quantum dots.

\section{Theoretical formalism} \label{sec:model}

\subsection{Model}

Here, we introduce the theoretical formalism for computing the level structure, or eigenenergies $E_l$ and dipole-allowed transitions between eigenstates $l$ and $m$ indicated by non-zero transition dipole moment $|\bm{d}_{lm}|$, of $N$ $M$-level emitters interacting \textit{via} dipole-dipole and hybridization interactions. For simplicity, we set $M=2$ throughout, 
\textit{i.e.} we study ensembles of two-level emitters, although this method can be straightforwardly generalized for any $M$. 

Each emitter $i$ is comprised of a ground orbital $|g_i\rangle$ and an excited orbital $|e_i\rangle$ with energy $\hbar\omega_i$, transition dipole moment $\bm{d}_i=\langle g_i | \mathrm{e} \hat{\bm{r}} | e_i \rangle$, and position $\bm{r}_i$, where $\hbar$ is the reduced Planck constant, e is the electron charge and $\hat{\bm{r}}$ is the position operator. We assume each orbital can be occupied by one electron. Therefore, in the number basis $|N^g_i, N^e_i\rangle$, there are four possible states: $|0_i^g, 0_i^e\rangle$, $|0_i^g, 1_i^e\rangle$, $|1_i^g, 0_i^e\rangle$, and $|1_i^g, 1_i^e\rangle$. 

The total Hamiltonian $\hat{H}$, including dipole-dipole and hybridization interactions, can be written as:
\begin{equation}
    \hat{H} = \hat{H}_0 + \hat{H}_\mathrm{dip} + \hat{H}_\mathrm{hyb},
\end{equation}
where the Hamiltonians $\hat{H}_0$, $\hat{H}_\mathrm{dip}$, and $\hat{H}_\mathrm{hyb}$ are for the bare emitters, dipole-dipole interaction, and hybridization interaction, respectively. The bare-emitter Hamiltonian can simply be written as $\hat{H}_0=\sum_i^N \hat{H}_i$, where the isolated emitter Hamiltonian $\hat{H}_i = \hbar \omega_i \hat{a}^\dag_{e,i} \hat{a}_{e,i}$, and $\hat{a}^\dag_{o, i}$ ($\hat{a}_{o, i}$) is the creation (annihilation) operator for an electron in orbital $o \in \{g, e\}$ of emitter $i$.

The dipole-dipole interaction Hamiltonian is
\begin{equation}
    \hat{H}_\mathrm{dip} = \sum_{i, j>i}^N J_{ij} \hat{\bm{d}}_i \hat{\bm{d}}_j,
\end{equation}
where $\hat{\bm{d}}_i = \bm{d}_i(\hat{a}^\dag_{e,i} \hat{a}_{g,i}+\hat{a}^\dag_{g,i} \hat{a}_{e,i})$ is the dipole operator. Notably, we retain all double (de-)excitations to preclude limitations of the rotating wave approximation. $J_{ij}$ is the dipole interaction energy given by
\begin{equation} \label{eq:Jij}
    J_{ij} = \frac{|\bm{d}_{i} ||\bm{d}_{j}| |}{4\pi\epsilon_0\epsilon_r|\bm{r}_i-\bm{r}_j|^3}\left[\bm{n}_{i}\cdot\bm{n}_{j}-3(\bm{n}_{i}\cdot\bm{n}_{ij})(\bm{n}_{j}\cdot\bm{n}_{ij})\right],
\end{equation}
where $\epsilon_r$ is the relative permittivity, $\bm{n}_{i}$ is the unit vector of the dipole moment $\bm{d}_{i}$, and $\bm{n}_{ij}$ is the unit vector of $\bm{r}_i-\bm{r}_j$. Notably, this form of the dipole-dipole interaction is appropriate only when the dipole-dipole distance $|\mathbf{r}_i - \mathbf{r}_j|$ is smaller than the transition wavelength $\lambda$ \cite{Ficek2002}. States that are dark under this approximation can, in fact, emit radiation with the full form of the dipole-dipole interaction. We quantitatively estimate the impact of this approximation for the parameters studied here in Appendix \ref{app:dipoleinteraction}.

The hybridization interaction Hamiltonian is
\begin{equation}
    \hat{H}_\mathrm{hyb} = \sum_{i,j>i}^N G^\mathrm{e}_{ij} (\hat{a}^\dag_{e,i} \hat{a}_{e,j} + 
    \hat{a}^\dag_{e,j} \hat{a}_{e,i}).
\end{equation}
where $G^\mathrm{e}_{ij}$ is the hybridization interaction energy between excited orbitals $|e_i\rangle$ and $|e_j\rangle$. This interaction is akin to inter-emitter electron, or charge, transfer that has been well studied in the molecular aggregates community \cite{Hestand2018}. Note that we ignore hybridization interactions between orbitals $|e_i\rangle$ and $| g_j \rangle$ because they are distant in energy, as well as interactions between ground orbitals $|g_i\rangle$ and $|g_j\rangle$ because the ground orbitals of real emitters of interest, such as excitons in quantum dots, are typically tightly localized. In principle, however, one can trivially add, for instance, the ground orbital hybridization term  $\sum_{i,j>i}^N G^g_{ij} (\hat{a}^\dag_{g,i} \hat{a}_{g,j} + \hat{a}^\dag_{g,j} \hat{a}_{g,i})$ to $\hat{H}_\mathrm{hyb}$.

Finally, we determine the level diagram of $N$ two-level emitters by diagonalizing $H$ and transforming the total dipole operator $\hat{\bm{d}}=\sum_i \hat{\bm{d}}_i$ into the eigenbasis. With the level diagram, we can evaluate the quality, specifically the efficiency $\eta$ and fidelity $\mathcal{F}$, of the emitted photons \textit{via} cascade decay from multiply excited states of cascade emitters, as described in Appendix \ref{app:evaluate}.

\section{Entanglement}

\subsection{Asymmetric mixing for Bell states}

\begin{figure}[!tbhp]
\centering
\includegraphics[angle=0,width=\linewidth]{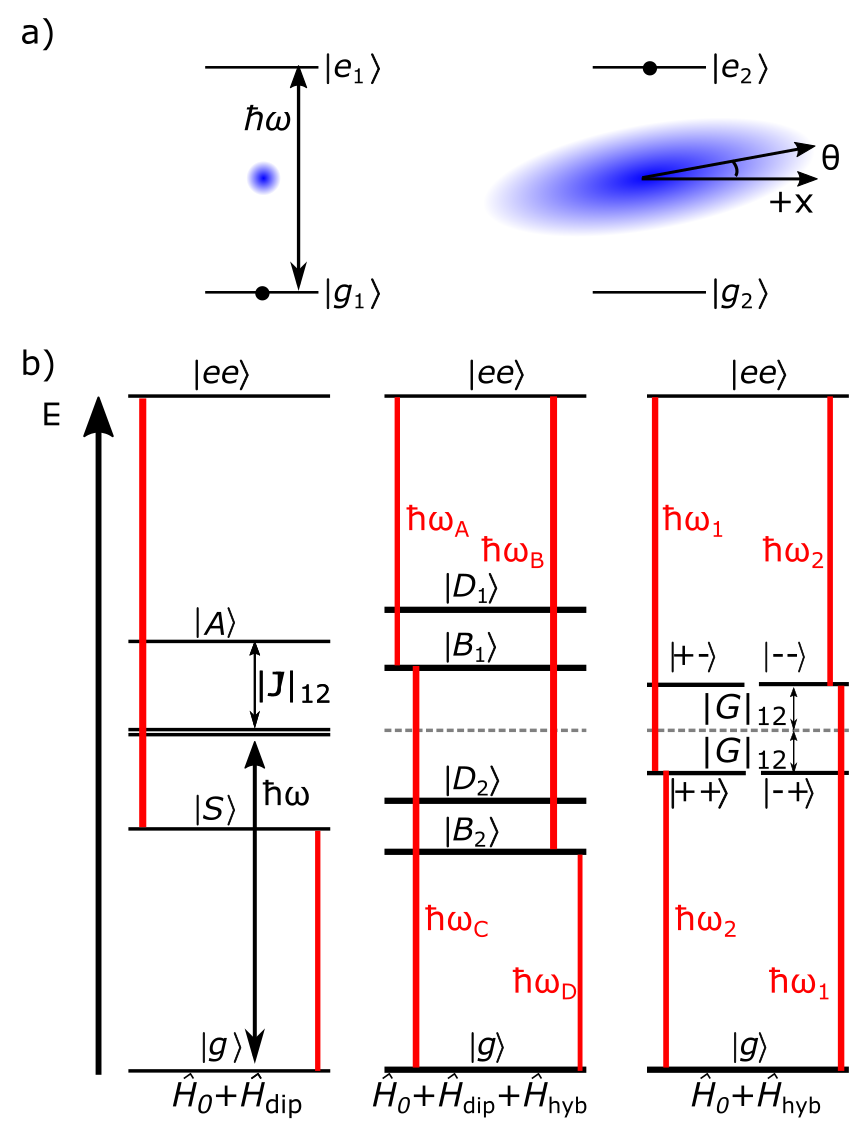}
\caption{a) Schematic of $N=2$ two-level emitters on the $x$-axis with transition dipole moments $\bm{d}_1=d_x \hat{i}$ and $\bm{d}_2=d_x [\mathrm{cos}(\theta)\hat{i}+\mathrm{sin}(\theta)\hat{j}]$. We assume the wave functions of the ground state are tightly localized, while the excited state are delocalized enough to interact through the hybridization interaction. b) Level diagrams for $N=2$ two-level emitters for $\hat{H}=\hat{H}_0+\hat{H}_\mathrm{dip}$ (left), $\hat{H}=\hat{H}_0+\hat{H}_\mathrm{dip}+\hat{H}_\mathrm{hyb}$ (middle), and  $\hat{H}=\hat{H}_0+\hat{H}_\mathrm{hyb}$ (right). Allowed transitions, or those with non-zero transition dipole moments, \textit{via} cascade decay from the doubly excited state $|ee\rangle$ are in red. Emission of frequency-entangled Bell states is only possible for two two-level emitters with both dipole-dipole and hybridization interactions.
}
\label{fig:bell}
\end{figure}

We explain why the combination of both dipole-dipole and hybridization interactions enable fuller control over the level diagram of $N$ two-level emitters, including those that are amenable to the emission of entangled photons. As an example, we study in detail the simple scenario illustrated in Fig. \ref{fig:bell}(a) where there are two emitters ($N=2$) on the $x$-axis ($\bm{r}_1-\bm{r}_2=r_x \hat{i}$) with identical orbital energies $\hbar\omega = \hbar\omega_1 = \hbar\omega_2$. We also assume $\theta$, or the angle between the $x$-axis and the transition dipole moment of the second emitter, is set to 0 so that both have $x$-polarized transition dipole moments $\bm{d}_1 = \bm{d}_2 = d_x \hat{i}$. Finally, we assume that each emitter contributes one electron for a total two electrons, giving 6 possible states in the number basis $|N^g_1, N^e_1, N^g_2, N^e_2\rangle$: $|0_1^g, 0_1^e, 1_2^g, 1_2^e\rangle$, $|0_1^g, 1_1^e, 0_2^g, 1_2^e\rangle$, $|0_1^g, 1_1^e, 1_2^g, 0_2^e\rangle$, $|1_1^g, 0_1^e, 0_2^g, 1_2^e\rangle$, $|1_1^g, 0_1^e, 1_2^g, 0_2^e\rangle$, and $|1_1^g, 1_1^e, 0_2^g, 0_2^e\rangle$. For notational convenience, we label these number states as, \textit{e.g.}, $|0_1^g, 0_1^e, 1_2^g, 1_2^e\rangle \equiv |g_2 e_2\rangle$, where only the occupied orbitals are included.

First, to understand the role of dipole-dipole coupling, we plot the level diagram without hybridization ($\hat{H}=\hat{H}_0+\hat{H}_\mathrm{dip}$) on the left of Fig. \ref{fig:bell}. In this well known result, the eigenstates $|l\rangle$ and eigenenergies $E_l$, in order of increasing $E_l$, are as follows: ground state $|g\rangle \approx |g_1 g_2 \rangle$ with $E_g\approx 0$; symmetric bright state $|S\rangle \approx 1/\sqrt{2}(|e_1 g_2\rangle+|g_1 e_2 \rangle)$ with $E_S \approx \hbar\omega - |J_{12}|$; two states $|g_2 e_2\rangle$ and $|g_1 e_1\rangle$ corresponding to double occupation of emitter 1 and 2, respectively, both with energy of $\hbar\omega$; anti-symmetric dark state $|A\rangle \approx 1/\sqrt{2}(|e_1 g_2 \rangle-|g_2 e_2 \rangle)$ with $E_A\approx \hbar\omega + |J_{12}|$; and doubly excited state $|ee\rangle \approx |e_1 e_2\rangle$ with $E_{ee} \approx 2\hbar\omega$. Note that the energies $E_l$ are generally listed here with their approximate values as opposed to exact ones due to the inclusion of the double (de-)excitation terms. From these eigenstates, we see that the dipole coupling mixes two of the singly excited states $|e_1 g_2\rangle$ and $|g_1 e_2\rangle$ with each other, leaving the other two $|g_1 e_1\rangle$ and $|g_2 e_2\rangle$ untouched. Assuming that the doubly excited state $|ee\rangle$ is initialized with a population of 1, the only dipole-allowed cascade decay path is $|ee\rangle \rightarrow |S\rangle \rightarrow |g\rangle$. While such a cascade decay would emit two photons with energies $E_{ee}-E_S \approx \hbar\omega + |J_{12}|$ and $E_S - E_g \approx \hbar\omega - |J_{12}|$, they are not entangled.

Now we seek understanding of the role of the hybridization interaction in the level diagram by plotting the level diagram without dipole-dipole coupling ($\hat{H}=\hat{H}_0+\hat{H}_\mathrm{hyb}$) on the right of Fig. \ref{fig:bell}. Again, in order of increasing energy $E_l$, the eigenstates $|l\rangle$ and eigenenergies $E_l$ are as follows when $G_{12}<0$: ground state $|g\rangle = |g_1 g_2\rangle$ with $E_g = 0$; $|++\rangle = 1/2(|g_1\rangle + |g_2\rangle)(|e_1\rangle + |e_2\rangle)$ and $|-+\rangle = 1/2(|g_1\rangle - |g_2\rangle)(|e_1\rangle + |e_2\rangle)$ with $E_{++} = E_{-+} = \hbar\omega-|G_{12}|$; two singly excited states $|+-\rangle = 1/2(|g_1\rangle + |g_2\rangle)(|e_1\rangle - |e_2\rangle)$ and $|--\rangle = 1/2(|g_1\rangle - |g_2\rangle)(|e_1\rangle - |e_2\rangle)$ with $E_{--} = E_{+-} = \hbar\omega+|G_{12}|$; and doubly excited state $|ee\rangle = |e_1 e_2\rangle$ with $E_{ee} = 2\hbar\omega$. From these eigenstates, we see that the hybridization interaction mixes all four of the singly excited states, as opposed to the dipole-dipole coupling interaction that only mixes two of the four. In this case, there are two possible decay paths. 

At first blush this cascade decay may seem appropriate for emission of entangled photons if the logical basis states $|0_1^L\rangle$ and $|1_1^L\rangle$  of the first emitted photon are assigned to the photon with energy $\hbar\omega_1=E_{ee}-E_{++}$ and $\hbar\omega_2=E_{ee}-E_{--}$, respectively, and the logical basis states $|0_2^L\rangle$ and $|1_2^L\rangle$  of the second emitted photon are assigned to the photon with energy $\hbar\omega_3=E_{++}-E_g$ and $\hbar\omega_4=E_{--}-E_g$. However, due to the equally weighted mixing between all four singly excited states, $\hbar\omega_1=\hbar\omega_4$ and $\hbar\omega_2=\hbar\omega_3$. Therefore, the two photons emitted by the pathway on the left are overall the same as the photons emitted on the right, resulting in zero entanglement. Finally, note that because $E_{--}=E_{+-}$ and $E_{++}=E_{-+}$, the choice of eigenvectors is arbitrary. For instance, the singly excited eigenstates $|1\rangle = 1/\sqrt{2}(|--\rangle + |+-\rangle)$ and $|2\rangle = 1/\sqrt{2}(|--\rangle - |+-\rangle)$ are equally valid states with energy $E_{--}=E_{+-}$, while the singly excited eigenstates $|3\rangle = 1/\sqrt{2}(|++\rangle + |-+\rangle)$ and $|4\rangle = 1/\sqrt{2}(|++\rangle - |-+\rangle)$ are equally valid states with energy $E_{++}=E_{-+}$. In this particular case there would only be four decay pathways from $|ee\rangle$ to each of the singly excited eigenstates to $|gg\rangle$. Despite this change in basis, the result is the same: an entangled photon pair would not be emitted. 

To achieve emission of frequency-entangled photon pairs, we require another interaction to asymmetrically shift the energies of the bright eigenstates of $\hat{H}=\hat{H}_0+\hat{H}_\mathrm{hyb}$. Adding the dipole-dipole coupling interaction, which mixes only two of the singly excited states $|e_1 g_2\rangle$ and $|g_1 e_2\rangle$, as opposed to all four in the case of only hybridization, can asymmetrically shift the energies. We plot the resulting level diagram for the simple scenario with both dipole-dipole coupling and hybridization in the center of Fig. \ref{fig:bell}. Because the eigenvectors and eigenenergies of the singly excited states in the basis of the states of isolated emitters are no longer analytically simple, we simply label the eigenstates as $|g\rangle$, bright states $|B_1\rangle$ and $|B_2\rangle$, dark states $|D_1\rangle$ and $|D_2\rangle$, and $|ee\rangle$. The allowed transitions $|ee\rangle \rightarrow |B_1\rangle \rightarrow |g\rangle$ and $|ee\rangle \rightarrow |B_2\rangle \rightarrow |g\rangle$ all emit photons with unique frequencies from each other. Assigning logical basis states $|0_1^L\rangle$ and $|1_1^L\rangle$  of the first emitted photon to $\hbar\omega_A$ and $\hbar\omega_B$ and $|0_2^L\rangle$ and $|1_2^L\rangle$  of the second emitted photon to $\hbar\omega_C$ and $\hbar\omega_D$, we qualitatively see that a Bell-like state $\alpha|0_1^L 0_2^L\rangle +\beta|1_1^L 1_2^L\rangle)$ can be emitted, where $\alpha$ and $\beta$ depend on the relative weights of each decay path and accrued phases.

\subsection{Entanglement optimization}

\begin{figure}[!tbhp]
\centering
\includegraphics[angle=0,width=\linewidth]{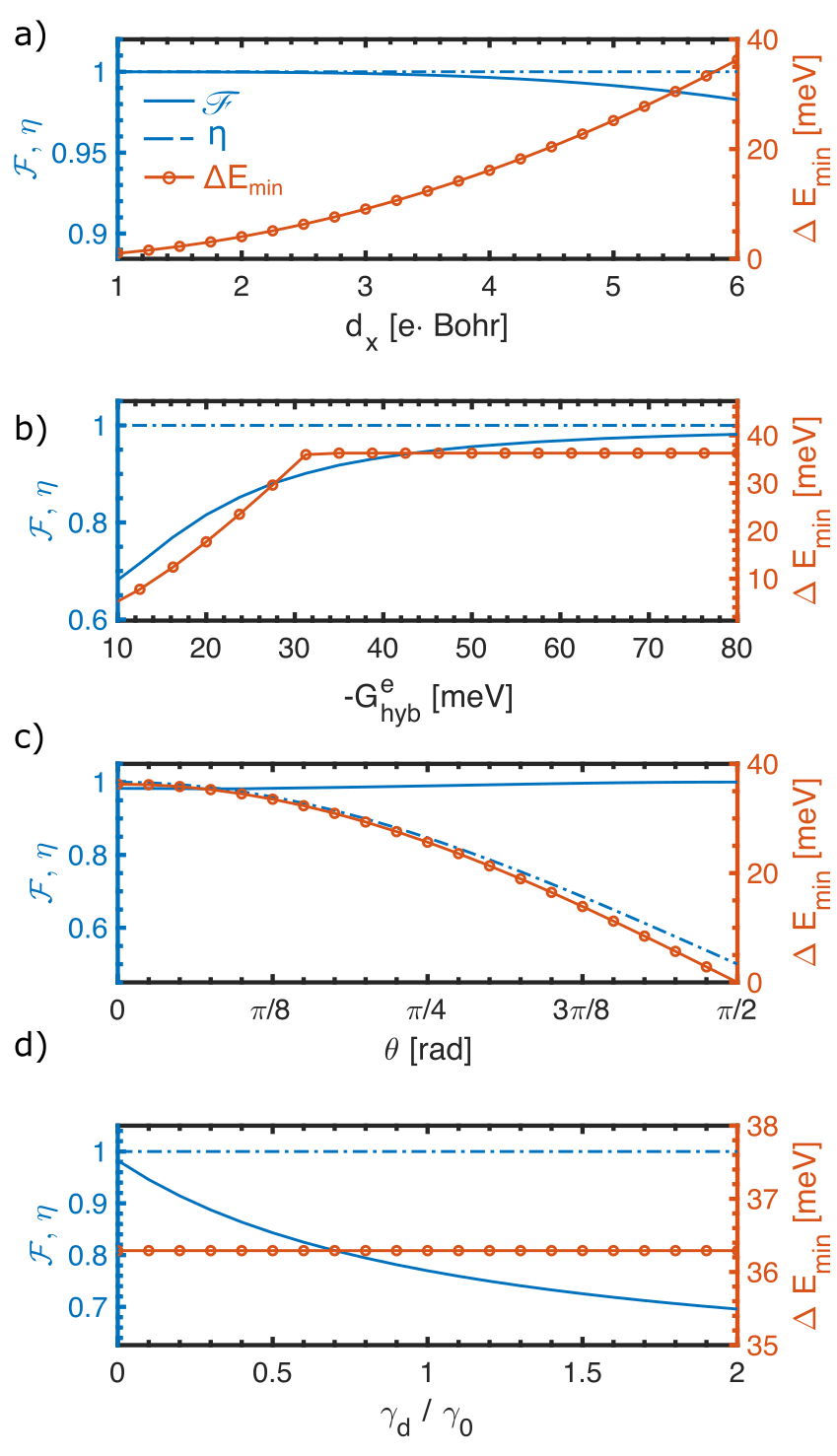}
\caption{We sweep (a) the magnitude of the transition dipole moment $d_x$, (b) the excited state hybridization interaction energy $G^e_\mathrm{hyb}$, (c) the relative angle $\theta$ between the transition dipole moment vectors $\bm{d}_1$ and $\bm{d}_2$, and (d) the dephasing rate $\gamma_\mathrm{d}$ relative to the fastest radiative decay rate $\mathrm{max}(\gamma_r^{lm})$. In each subplot, we show the fidelity $\mathcal{F}$ (solid blue) with an ideal Bell state, efficiency $\eta$ (dotted blue), and $\Delta E_\mathrm{min}$ (solid orange), or the minimum energy difference between emitted photons mapped onto logical basis states to determine the maximal photon peak broadening permissible for frequency resolution. All system parameters not being sweep in each respective plot are as follows: $\hbar\omega = 1$ eV, $d_x=|\bm{d}_1|=|\bm{d}_2|=6$ e$\cdot$Bohr, $\bm{r}_1-\bm{r}_2 = 40\hat{i}$ in Bohr, $G^e_\mathrm{hyb}=80$ meV, and $\epsilon_r=1$.  Increasing the magnitude of the hybridization interaction increases $\mathcal{F}$ and $\Delta E_\mathrm{min}$. Meanwhile, increasing the dipole-dipole interaction \textit{via} increasing $d_x$ decreases $\mathcal{F}$ but eases the challenge of resolving the photon frequencies due to increasing $\Delta E_\mathrm{min}$. $\mathcal{F}$ and $\Delta E_\mathrm{min}$ are stable for small deviations around $\theta=0$ but drop precipitously near $\theta=\pi/2$ at which point the dipole-dipole interaction disappears. Decoherence rates on the order of and higher than the emission rate reduces entanglement fidelity.
}
\label{fig:belloptimize}
\end{figure}

We optimize the entanglement by tuning the system parameters, specifically the excited state hybridization interaction energy $G^\mathrm{e}_\mathrm{hyb}$, transition dipole moment magnitude $d_x$, relative angle $\theta$ between the transition dipole moment vectors $\bm{d}_1$ and $\bm{d}_2$, and dephasing rate $\gamma_\mathrm{d}$, and plot the results in Fig. \ref{fig:belloptimize}(a)-(d), respectively. Evaluating the level diagrams for cascade emission of entangled photons as discussed in Appendix \ref{app:evaluate}, we efficiently determine the impact of these parameters on fidelity $\mathcal{F}$ with and efficiency $\eta$ of emitting an ideal Bell state $|\phi_\mathrm{B}\rangle = 1/\sqrt{2}(|0_1^L 0_2^L\rangle +|1_1^L 1_2^L\rangle)$. In addition to $\mathcal{F}$ and $\eta$, we plot the minimum energy difference $\Delta E_\mathrm{min}$ between all emitted photons that are mapped onto the logical basis states. By doing so, we determine the maximum line broadening permissible to resolve photons from each other \textit{via} their frequencies.

We first sweep the magnitude of the transition dipole moment $d_x$ in Fig. \ref{fig:belloptimize}(a). Throughout this range, $\eta$ remains 1, suggesting that all of the population follows the two decay paths resulting in the two superpositioned states in the Bell state. As $d_x$ increases, $\Delta E_\mathrm{min}$ increases to $\sim$36 meV at $d_x=6$ e$\cdot$Bohr, while $\mathcal{F}$ decreases from 1 to 0.97; the former can be understood as a result of the increasing dipole-dipole interaction energy, while the latter is a result of the two decay paths having increasingly different magnitudes of weights.

In Fig. \ref{fig:belloptimize}(b), we then analyze the effect of changing the hybridization energy $G_\mathrm{hyb}^\mathrm{e}$. Again, throughout this range, $\eta$ remains 1. However, in this case, with increasing hybridization interaction energy, the fidelity $\mathcal{F}$ increases, while the $\Delta E_\mathrm{min}$ increases until $\sim$36 meV. $\mathcal{F}$ increases with increasing magnitude of $G^e_\mathrm{hyb}$ for a fixed dipole-dipole interaction because the weights of the two decay paths equalize, highlighting the importance of the presence of the hybridization interaction in addition to the dipole-dipole coupling for high fidelity emission of entangled photon pairs. $\Delta E_\mathrm{min}$ saturates at $\sim$36 meV with increasing magnitude of $G_\mathrm{hyb}^e$ because at that point, it is limited by the dipole-dipole interaction---recall that at $d_x=6$ e$\cdot$Bohr in Fig. \ref{fig:belloptimize}(a), the same value of $d_x$ in Fig. \ref{fig:belloptimize}(b), $\Delta E_\mathrm{min}\sim36$ meV.

We sweep the angle $\theta$ between the transition dipole moments of emitters 1 and 2 in Fig. \ref{fig:belloptimize}(c). Here, while $\mathcal{F}$ remains close to 1, both $\Delta E_\mathrm{min}$ and $\eta$ drop precipitously near $\theta=\pi/2$. At this value, the dipole-dipole coupling is 0, resulting in the effective Hamiltonian $\hat{H}=\hat{H}_0+\hat{H}_\mathrm{hyb}$ discussed previously and whose level structure is plotted on the right side of Fig. \ref{fig:bell}(b).

Finally, in Fig. \ref{fig:belloptimize}(d), we plot the effect of dephasing $\gamma_\mathrm{d}$ as a proportion of the bare emitter decay rate $\gamma_0=\mathcal{C}d_x^2$ where $d_x=6$ e$\cdot$Bohr. Both $\eta$ and $\Delta E_\mathrm{min}$ remain constant, as expected because pure decoherence should not change the central emission frequency nor result in population loss. The fidelity $\mathcal{F}$, meanwhile, is quite sensitive to $\gamma_\mathrm{d}$, validating the physical intuition that high-quality quantum emission requires fast radiative decay rates relative to environment loss and decoherence.

\subsection{Arbitrarily entangled photons: GHZ states}

\begin{figure}[!tbhp]
\centering
\includegraphics[angle=0,width=1.0\linewidth]{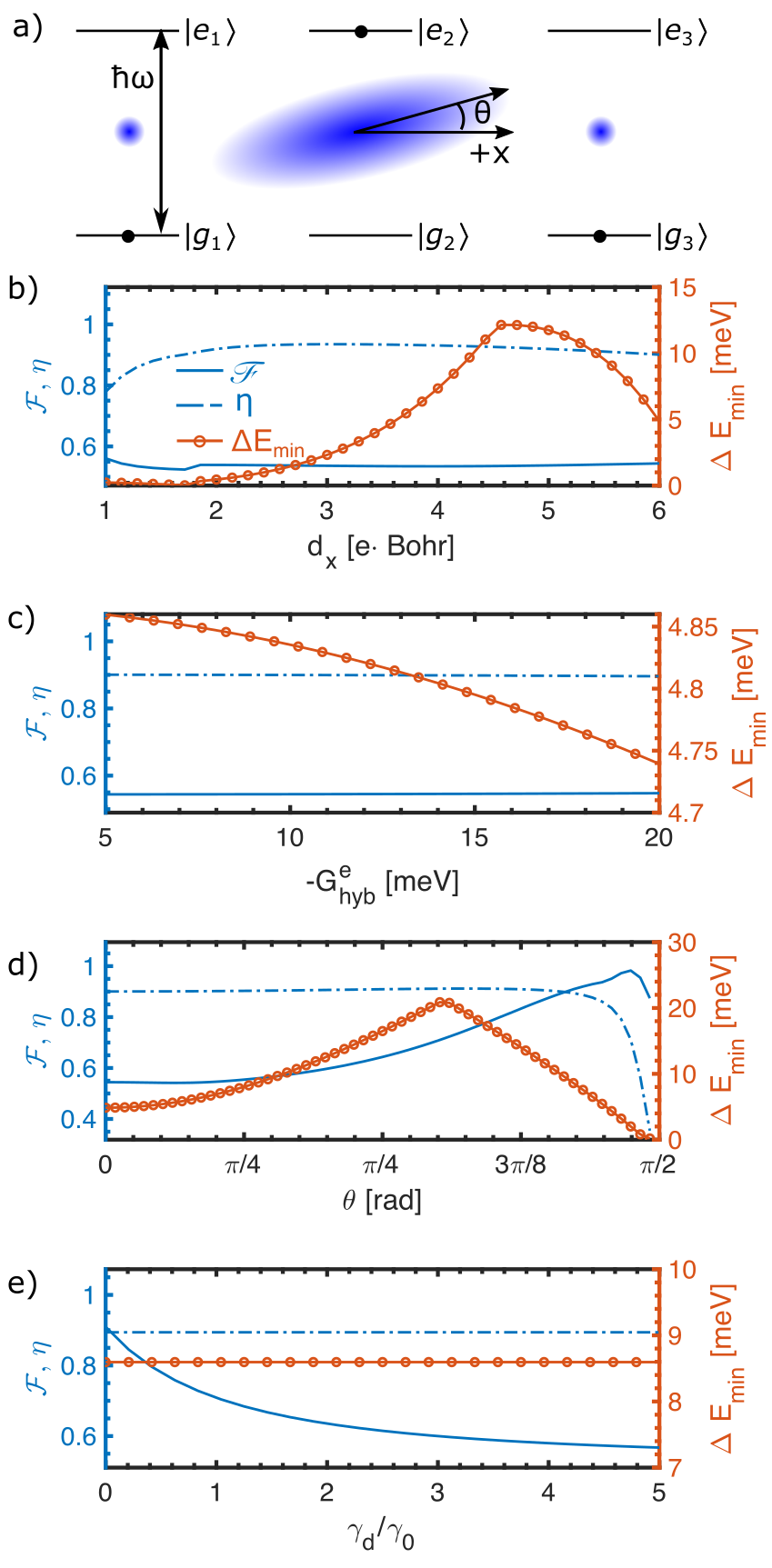}
\caption{(a) Configuration for $N=3$ two-level emitters positioned on a line parallel to their dipole moments $\bm{d}_i$ and capable of emitting frequency-entangled GHZ states \textit{via} cascade decay from the triply excited state $|20\rangle$. We sweep (b) $G^\mathrm{e}_\mathrm{hyb}$, (c) $d_x$, (d) $\theta$, and (e) $\gamma_\mathrm{d}$. All system parameters not being sweep in each respective plot are assumed to be as follows: $\hbar\omega = 1$ eV, $d_x=|\bm{d}_1|=|\bm{d}_2|=|d_3|=6$ e$\cdot$Bohr, $\bm{r}_3-\bm{r}_2=\bm{r}_1-\bm{r}_2 = 20\hat{i}$ in Bohr except in (e) where $\theta$ is set to the value that maximizes $\mathcal{F}$ in (d), $G^\mathrm{e}_\mathrm{hyb}=80$ meV, and $\epsilon_r=1$.  By optimizing the system parameters, $\mathcal{F}$ and $\eta$ can be simultaneously 90\% with $\Delta E_\mathrm{min}\sim10$ meV.
}
\label{fig:GHZ}
\end{figure}

This method of constructing new level diagrams of composite emitters by coupling individual emitters \textit{via} hybridization and dipole-dipole interactions can be generalized to construct level structures amenable to emission of arbitrarily entangled photons. As an example, we optimize the configuration in Fig. \ref{fig:GHZ}(a) for emission of a three-photon GHZ state $1/\sqrt{2}(|000\rangle + |111\rangle)$, where the labels for the logical basis have been dropped for convenience. All three two-level emitters lie on the $x$-axis with equal magnitude dipole moments that we assume to be in the $x$-direction for the left and right emitters, while the transition dipole moment vector of the middle emitter can be rotated $\theta$ from the $x$-axis. In this system, there can be $>10^5$ possible decay paths.

We again sweep the excited state hybridization interaction energy $G^\mathrm{e}_\mathrm{hyb}$, transition dipole moment magnitude $d_x$, relative angle $\theta$ between the transition dipole moment vectors $\bm{d}_1$ and $\bm{d}_2$, and dephasing rate $\gamma_\mathrm{d}$, and plot the impact on $\mathcal{F}$, $\eta$, and $\Delta E_\mathrm{min}$ in Fig. \ref{fig:GHZ}(b)-(e), respectively. 
In Fig. \ref{fig:GHZ}(b) and (c), we see that the maximum $\eta$ is only around 0.9 with a significantly lower $\mathcal{F}=0.55$. While the quality of entanglement is low, interestingly, all entanglement measures are much more stable with respect to changes in both $d_x$ and $G^\mathrm{e}_\mathrm{hyb}$ compared to the Bell state structures studied in Fig. \ref{fig:belloptimize}. For instance, in (c), $\mathcal{F}$ and $\eta$ change by less than 1\% and $\Delta E_\mathrm{min}$ is stable within 0.1 meV. Stronger performance is possible by tuning $\theta$, where between $\theta=\pi/4$ and $\pi/2$, both $\mathcal{F}$ and $\eta$ can be nearly 0.90 with $\Delta E_\mathrm{min}$ approaching 10 meV, or $\mathcal{F}$ can be as high as 0.97 with lower $\eta=0.70$. While we expect further improvements to be possible with multi-dimensional optimization techniques, these results already demonstrate the tantalizing promise of high-efficiency, high-fidelity, deterministic emission of arbitrary entangled photon states.

\section{Conclusions and outlook}
In summary, we leverage hybridization and dipole-dipole interactions between simple emitters to construct composite emitters. These composite emitters have particularly valuable applications as deterministic sources of entangled photon states \textit{via} cascade decay from multiply excited states. As a simple example, we study two two-level emitters for emission of Bell states. We explain why the combination of both types of interactions is necessary for emission of Bell states. We also explain why varying system parameters, including the strength of the hybridization interaction $G^\mathrm{e}_\mathrm{hyb}$ and direction and magnitude of the transition dipole moment $\bm{d}_{lm}$, affects three relevant metrics: fidelity $\mathcal{F}$, efficiency $\eta$, and minimum energy difference $\Delta E_\mathrm{min}$ between emitted photons, where the latter is necessary to predict the maximum possible line width broadening for successful resolution of detected photon frequencies. Finally, we demonstrate how cascade emitters of arbitrarily entangled photons can be rationally designed by optimizing, as an example, a composite emitter for GHZ states, a more complex computational problem due to the many possible decay paths and the many more states. By simply sweeping the system parameters, we achieve a fidelity $\mathcal{F}$ and efficiency $\eta$ as high as 90\% with $\Delta E_\mathrm{min}$ on the order of meV. 

\textit{Ab initio} computation of eigenstates of, eigenenergies of, and transtions between multiply excited state is a challenging problem \cite{Loos2019}, thereby highlighting the major benefit of this toy model-driven approach to guiding rational design of composite emitters. The present method can accelerate optimization of composite emitters for applications with many constraints on frequency-entangled photons, such as quantum networking \cite{Chen2016, Pan2012, Sangouard2011}, where optical, mid-infrared, and microwave photons are relevant for on-chip computation, transmission across long distances, and coupling to other qubit types \cite{Wang2021Perspective}, such as superconducting qubits, respectively.

We anticipate several fruitful research directions toward designer cascade emitters from real emitters, such as moir\'e excitons, molecular aggregates, and quantum dots. Each of these emitter types can be bright sources of single photons \cite{Baek2020, Lin2017}, implying that these emitters can have relatively large transition dipole moments resulting in strong dipole-diple coupling and fast radiative decay, as well as relatively low loss and decoherence rates. In addition, unlike in neutral atoms and defects in solid-state materials, each of these systems can exhibit hybridization interactions: in moir\'e excitons, the extent of their wave functions can be comparatively large and give rise to hybridization-like interactions: changing the layer-layer rotation angle can change the depth of the moir\'e potential \cite{Tran2019} and delocalize the exciton wave functions; molecular aggregates can be placed close enough together to hybridize; and colloidal quantum dots have exhibited signatures of direct coupling \cite{Cui2019}. We believe that molecular aggregates \cite{Hestand2018} and colloidal quantum dots are especially strong candidates for realization of this approach. Colloidal quantum dots, in particular, have exhibited both hybridization \cite{Cui2019} and dipole-dipole interactions \cite{Philbin2021} on the order of tens of meV, and the optical properties of single nanoparticles can be straightforwardly computed from first principles. When investigating the application of our design approach to these physical systems, researchers should take particular note of effects of hybridization and dipole-dipole coupling on loss and decoherence channels, as has been extensively studied for atoms \cite{Lukyanets2006}. We also note that pumping schemes that achieve efficient population of the multiply excited state must be designed with care, where the optimal pumping scheme can be unique to emitter type. There are, for instance, a multitude of pumping schemes for biexcitons in semiconductor quantum dots \cite{Carmele2019, Forstner2003, Nazir2016, Denning2020, Poddubny2012pump, Bauch2021}.

\section*{Associated content}
Code to reproduce the calculations in this paper are available at \href{https://github.com/drekwang/cascadeDecay}{https://github.com/drekwang/cascadeDecay}.

\section*{Acknowledgements}
D.S.W. and I.A. contributed equally to this work. The authors acknowledge valuable discussions with John Philbin, Stefan Ostermann, Valentin Walther, and Tom\'{a}\v{s} Neuman. D.S.W. is an NSF Graduate Research Fellow. S.F.Y. would like to acknowledge funding by DOE, AFOSR, and NSF.

\appendix

\section{Dipole-dipole interaction} \label{app:dipoleinteraction}
The form of the dipole-dipole interaction in Eq. \eqref{eq:Jij} is valid only when the inter-emitter distance $|\mathbf{r}_i - \mathbf{r}_j|=r_{ij}$ is much smaller than the bare-atom transition wavelength $\lambda_0$, or $r_{ij}/\lambda_0=\xi \ll 1$ \cite{Ficek2002}. For concreteness, we quantitatively estimate the impact of this approximation on the results for the case of two two-level emitters, as in Fig. \ref{fig:bell} and \ref{fig:belloptimize}, noting that similar arguments can be made for more complex systems, such as the case of three two-level emitters in Fig. \ref{fig:GHZ}. We adapt the notation of Ref.~\citenum{LaurinOstermann2016}.

Under the full expression for the dipole-dipole interaction, the anti-symmetric dark state can, in fact, emit photons with rate $\gamma_\mathrm{A}=\Gamma[1-F(\xi)]$, while the symmetric bright state with emission rate $\gamma_\mathrm{S}=\Gamma[1+F(\xi)]$ is less bright than under the approximated form, where $\Gamma=\mathcal{C}|\mathbf{d}_1|^2=\mathcal{C}|\mathbf{d}_2|^2$ is the bare-emitter radiative decay rate, and $F(\xi)$ is the correction factor defined as
\begin{subequations}
\begin{align}
    F(\xi) &= \frac{3}{2} \big[(1-\mathrm{cos}^2\theta)\frac{\mathrm{sin}\xi}{\xi} \\
    &+(1-3\mathrm{cos}^2\theta)(\frac{\mathrm{cos}\xi}{\xi^2}-\frac{\mathrm{sin}\xi}{\xi^3})\big].
\end{align}
\end{subequations}

In the case of two two-level emitters, $r_{ij}\sim$2 nm with bare-emitter transition wavelengths $\lambda_0\sim$1200 nm, giving $F(\xi)\sim 0.99999$ and implying that the entanglement measures $\mathcal{F}$ and $\eta$ are substantially more sensitive to the studied system parameters $d_x, G^\mathrm{e}_\mathrm{hyb}, \theta,$ and $\gamma_\mathrm{d}$ than the neglected emission from dark states.

\section{Level diagram evaluation} \label{app:evaluate}

We seek to calculate the quality, specifically the efficiency $\eta$ and fidelity $\mathcal{F}$, of the emitted photons \textit{via} cascade decay from multiply excited states of cascade emitters. Assuming the desired photon state is $|\phi\rangle = \sum_p^P |p\rangle$ with density matrix $\sigma=|\phi\rangle\langle\phi|$, where, \textit{e.g.}, $P=2$ for the Bell and GHZ states we study in further detail in this manuscript, then both measures can be computed from the emitted photon density matrix $\rho$ as \cite{Jozsa1994}
\begin{equation} \label{eq:eta}
    \eta = (\sum_p^P \rho_{pp}) / \mathrm{Tr}[\rho],
\end{equation}
\begin{equation}
    \mathcal{F}(\rho', \sigma) = \Big(\mathrm{Tr} \sqrt{\sqrt{\rho'} \sigma \sqrt{\rho'}}\Big)^2,
\end{equation}
where $\rho'=\rho / \sum_p \rho_{pp}$. (Note that because the computed $\rho$ is manually normalized in practice, we explicitly include $\mathrm{Tr}[\rho]$ in Equation \ref{eq:eta}.)  Therefore, the efficiency $\eta$ is the proportion of emitted photon states in one of the states $|p\rangle$ in the desired photon state $|\phi\rangle$, and the fidelity $\mathcal{F}$ is the overlap between the desired photon density matrix $\sigma$ among the emitted photon states in the states $|p\rangle$. 

We first show how to compute the full density matrix $\rho$ of the emitted photons.
Closely following a generalized version of the derivations shown in Refs. \citenum{Troiani2006, Pfanner2008}, a matrix element $\rho_{a,b}$ of the $P$-photon density matrix $\rho$ can be written as
\begin{equation}
\begin{aligned}
    \rho_{a,b} &= \mathrm{avg}[\langle  \hat{\sigma}^-_{\omega^a_1}(t_1)...\hat{\sigma}^-_{\omega^a_P}(t_P)\hat{\sigma}^+_{\omega^b_P}(t_P)...\hat{\sigma}^+_{\omega^b_1}(t_1)\rangle] \\
    & = \mathrm{avg}(\mathcal{G}_{a,b}),
\end{aligned}
\end{equation}
where the average is over all times $t_1 < ... < t_P$; $q \in \{a, b\}$ refers to an $P$-photon state created by cascade emission of photons with frequencies $\omega^q_1$, ..., $\omega^q_P$; $\hat{\sigma}^+_{\omega}(t)$ [$\hat{\sigma}^-_{\omega}(t)$] is the transition operator of the electronic transition $|l\rangle \rightarrow |m\rangle$ with frequency $\omega$ in the Heisenberg picture; and $\hat{\sigma}^+_{\omega}=\hat{a}^\dag_m \hat{a}_l$ [$\hat{\sigma}^-_{\omega}=\hat{a}^\dag_l \hat{a}_m$]. 

$\mathcal{G}_{a,b}$ can be computed from the dynamics of the electronic system undergoing cascade decay. The diagonalized Hamiltonian of the electronic system, in the absence of interaction with the environment and determined \textit{via} the procedure described in Section \ref{sec:model}, is
\begin{equation}
    \hat{H} = \sum_l E_l |l\rangle \langle l |,
\end{equation}
where the $l$th eigenstate $|l\rangle$ has energy $E_l$. The evolution of the density operator $\rho_\mathrm{el}$ of the electronic system can be described with a master equation of the Lindblad form \cite{scully_zubairy_1997}:
\begin{equation}
\begin{aligned}
    \mathrm{i} \dot{\rho}_\mathrm{el} &= \frac{1}{\hbar}[H, \rho_\mathrm{el}] - \frac{\mathrm{i}}{2}\sum_\mu (\hat{L}^\dag_\mu\hat{L}_\mu\rho_\mathrm{el} + \rho_\mathrm{el} \hat{L}^\dag_\mu\hat{L}_\mu-2\hat{L}_\mu\rho_\mathrm{el}\hat{L}^\dag_\mu) \\
    &= \mathcal{L}[\rho_\mathrm{el}],
\end{aligned}
\end{equation}
where the Lindblad operators $\hat{L}_\mu$ describe the interactions $\mu$ of the electronic system with the environment and $\mathcal{L}$ is the Liouville superoperator. As in Ref. \citenum{Pfanner2008}, we consider two main forms of interaction with the environment in quantum dots, the physical emitters we suggest for further study: radiative decay, which leads to emission of frequency-entangled photons, and pure dephasing from electron-phonon coupling and spectral diffusion. We write the former Lindblad operators as $\hat{L}_{\mathrm{r},lm}=\sqrt{\gamma^{lm}_r}|m\rangle \langle l|$. In agreement with Fermi's Golden Rule and Wigner-Weisskopf theory, we assume the radiative decay rate $\gamma^{lm}_r$ for transitions from higher-energy state $|l\rangle$ to lower-energy state $|m\rangle$ is proportional to $|\bm{d}_{lm}|^2$ and scaled by a constant $\mathcal{C}$. This constant $\mathcal{C}$ includes the photon density of states that, for simplicity, we assume to be constant for all photon frequencies, although this term could easily be generalized for any given cavity, waveguide, or free space configuration. For the dephasing process, we write the Lindblad operators as $\hat{L}_{\mathrm{d},l}=\sqrt{\gamma_d}|l\rangle \langle l|$. For simplicity, we assume the dephasing rate $\gamma_d$ is a constant, although this model could be straightforwardly generalized to describe the phenomenology of particular emitter systems. For instance, excitons in quantum dots or defects in solid-state materials both exhibit a zero-phonon line and a phonon tail that requires $\gamma_d$ to be described more microscopically and potentially in a non-Markovian manner \cite{Borri2001, Krummheuer2002, Jahnke2015}.

Using the quantum jump approach \cite{Plenio1998}, we can solve the master equation to find
\begin{equation}
    \rho_\mathrm{el}(t) = e^{-\mathrm{i}\mathcal{L}t}\rho_\mathrm{el}^0,
\end{equation}
where $\rho_\mathrm{el}^0$ is the initial density matrix at time $t=0$ and asssumed to be decoupled from the environment. Finally, with $\rho_\mathrm{el}(t)$, we can solve for $\mathcal{G}$ and, thus, the $N$-photon density matrix $\rho$ using the quantum regression theorem \cite{GardinerZoller2000, Guarnieri2017} and noting that the operator $A_j(t_j)$ is evolved in the Heisenberg picture as $A_j(t_j)=e^{+\mathrm{i}\mathcal{L}t_j} A_j e^{-\mathrm{i}\mathcal{L}t_j}$:
\begin{equation}
\begin{aligned}
    \mathcal{G}_{a,b} = \mathrm{Tr}\bigg[\hat{\sigma}^+_{\omega^b_P}(t_P)\Big[e^{-{\mathrm{i}}\mathcal{L}(t_P-t_{P-1})}...\big[\hat{\sigma}^+_{\omega^b_1}(t_1) \\ \times[e^{-{\mathrm{i}}\mathcal{L}(t_1)}\rho_\mathrm{el}^0] \hat{\sigma}^-_{\omega^a_1}(t_1)\big]...\Big]\hat{\sigma}^-_{\omega^a_P}(t_P)\bigg].
\end{aligned}
\end{equation}

It is possible to further adapt this calculation to experimental conditions by including, for instance, the efficiency of detection. It is also possible to improve the entanglement by spectrally filtering the output or delaying the detection time, as described further in Ref. \citenum{Pfanner2008}. However, these approaches are outside the scope of this study, which aims to focus on the level structures of the composite emitters themselves.

\begin{figure}[!tbhp]
\centering
\includegraphics[angle=0,width=\linewidth]{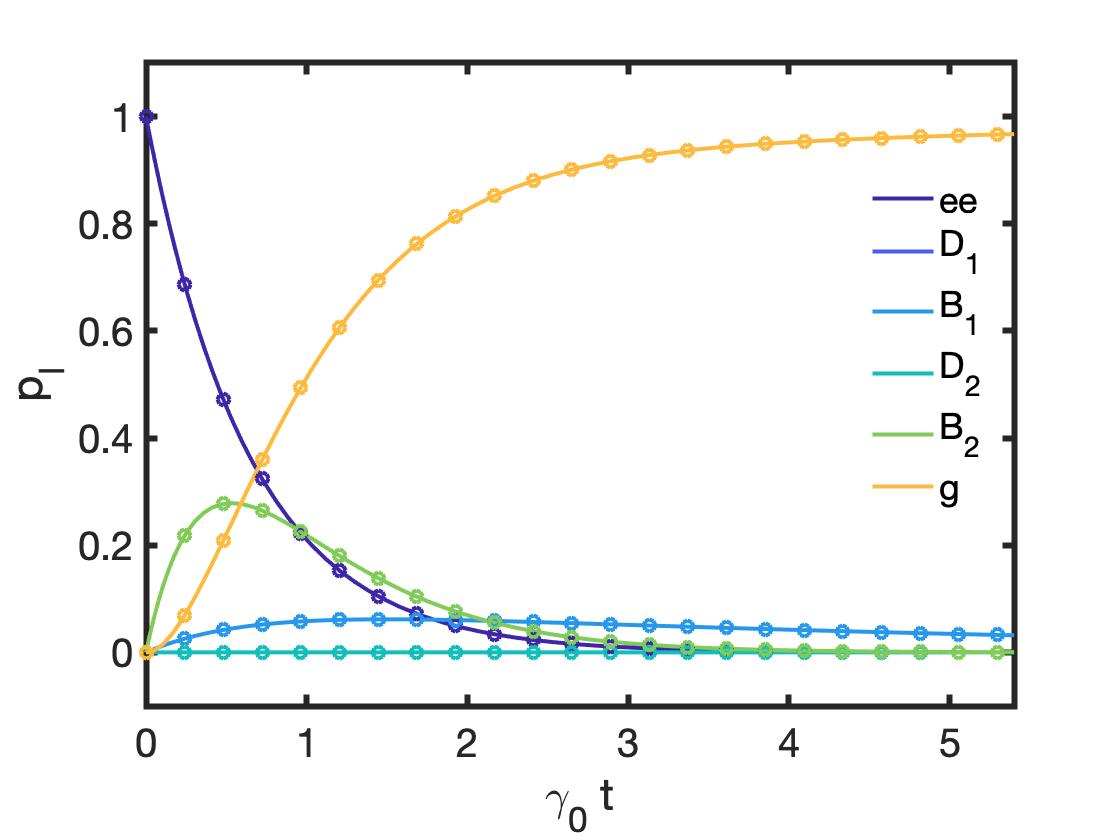}
\caption{Time-dependent population for cascade decay from the doubly excited state of two two-level systems using the quantum master equation (solid) and classical rate equation (circles) approaches. A schematic of such a level diagram is shown in the center of Fig. \ref{fig:bell}(b).  The system is initialized with $p_{ee}(t=0)=1$. We set $\hbar\omega = 1$ eV, $d_x=|\bm{d}_1|=|\bm{d}_2|=6$ e$\cdot$Bohr, $\bm{r}_1-\bm{r}_2 = 40\hat{i}$ in Bohr, $G^\mathrm{e}_\mathrm{hyb}=80$ meV, and $\epsilon_r=1$. The $x$-axis is unitless time, where the time $t$ is scaled by the radiative rate $\gamma_0 = \mathcal{C}|d_x|^2$ of the bare emitter. There is virtually zero difference in the time-dependent populations between the classical and quantum approaches in this case, enabling us to compute the on-diagonal terms of the photon density matrix $\rho$ more efficiently with the classical approach.
}
\label{fig:classicalvquantum}
\end{figure}

As computing the full density matrix $\rho$ can be resource-intensive, we seek a more efficient method of evaluating level diagrams. To this end, we observe that to compute $\eta$, we require only the on-diagonal elements of $\rho$, while $\mathcal{F}$ requires only the elements of $\rho$ of the $|i\rangle$ bases. Consider the computational cost-savings when analyzing level diagrams for cascade emission of, for instance, GHZ states from $N=3$ hybridized and dipole-coupled emitters. This system can have up to $\sim10^5$ decay paths or unique photon states and, therefore, a density matrix with $\sim10^5\times 10^5$ matrix elements, while we require just $\sim10^5$ matrix elements of $\rho$ for $\eta$ and only 4 elements are necessary for $\mathcal{F}$.

Computing all of the on-diagonal elements can still be an expensive process, however, as each requires a multidimensional integral over the product of several, potentially large matrices $\hat{\sigma}^\pm_\omega$ and linear maps  $\mathrm{exp}(-\mathrm{i}\mathcal{L}t)$. To more efficiently compute the on-diagonal elements of $\rho$, we turn to a classical rate equation approach. Given the diagonalized Hamiltonian and dipole operator in the eigenbasis, we determine all possible cascade decay pathways given some initial state, such as the $N$-excited state. With a general Runge-Kutta ordinary differential equation integrator, we propagate the rate equations $\diff{p_l}{t}=\sum_j(k_{lm}^\mathrm{in} p_m - k_{lm}^\mathrm{out} p_l)$, where rate constants $k_{lm}^\mathrm{in}$ ($k_{lm}^\mathrm{out}$) $= \mathcal{C} |\bm{d}_{lm}|^2$ for $E_l <$ ($>$) $E_m$, such that population transfer only from higher-energy to lower-energy states is allowed. As for the Lindbladian terms in the quantum master equation approach, both the scaling of $k_{lm}$ with $|\bm{d}_{lm}|$ and the transfer of energy from higher-energy to lower-energy states is generally expected from spontaneous emission into free space calculated \textit{via}, for instance, the Wigner-Weisskopf method or Fermi's Golden Rule, while $\mathcal{C}$ includes scaling due to the photon density of states that for simplicity we assume to be constant for all emitted photons. 

A comparison of the time-dependent population curves computed with the classical vs. quantum approaches for two two-level emitters capable of emitting Bell states is shown as an example in Fig. \ref{fig:classicalvquantum} for initial population of $p_{ee}(t=0)=1$. For both approaches, the population briefly transfers to the intermediate states $|B_1\rangle$ and $|B_2\rangle$ before eventually populating the ground state $|g\rangle$. Note that the dark states are never populated. The results agree closely, suggesting that the classical rate equation approach is appropriate for computing populations and, therefore, on-diagonal terms of the photon density matrix $\rho$.

From every state $l$, we then compute the relative outward flux $w_{lm}=\int k_{lm}^\mathrm{out} p_l \mathrm{d}t/\sum_m \int k_{lm}^\mathrm{out} p_l \mathrm{d} t$, where $\sum_m w_{lm} = 1$, of population from $l$ into states $m$, allowing us to compute the relative population transfer through the path comprising of transitions through states $l\rightarrow ... \rightarrow n$ as $w_l ... w_n$. This product then corresponds to the on-diagonal photon density matrix element $\rho_{l,...,n; l,...,n}$, or the population in the photon state created by decay pathway $l\rightarrow ... \rightarrow n$.

\newcommand{\noopsort}[1]{} \newcommand{\printfirst}[2]{#1}
  \newcommand{\singleletter}[1]{#1} \newcommand{\switchargs}[2]{#2#1}

\clearpage

\end{document}